\documentclass[a4paper,11pt]{article}

\usepackage{jinstpub} 
\usepackage{multirow}
\usepackage{placeins}
\usepackage{titlesec}
\begin{document}

\title{\boldmath Leveraging Staggered Tessellation for Enhanced Spatial Resolution in High-Granularity Calorimeters}
\author[a]{Sebouh J. Paul}
\author[a,b]{Miguel Arratia}
\date{\today}
\affiliation[a]{Department of Physics and Astronomy, University of California, Riverside, CA 92521, USA}
\affiliation[b]{Thomas Jefferson National Accelerator Facility, Newport News, Virginia 23606, USA}

\emailAdd{miguel.arratia@ucr.edu}
\abstract{We advance the concept of high-granularity calorimeters with staggered tessellations, underscoring the effectiveness of a design incorporating multifold staggering cycles based on hexagonal cells to enhance position resolution. Moreover, we introduce HEXPLIT, a sub-cell re-weighting algorithm tailored to harness staggered designs, resulting in additional performance improvements. By combining our proposed staggered design with HEXPLIT, we achieve an approximately twofold enhancement in position resolution for neutrons across a wide energy range, as compared to unstaggered designs. These findings hold the potential to elevate particle-flow performance across various forthcoming facilities.
}

\maketitle

\section{Introduction}

Highly-granular calorimeters, optimized for particle-flow applications~\cite{RevModPhys.88.015003,Thomson_2009}, offer unprecedented potential for the detailed measurement of particle showers. Hits in multiple layers can be used to determine the axis of a particle shower or the trajectory of a minimum-ionizing particle (MIP) with great accuracy.

When a MIP hits two cells with the same transverse position, the second hit adds very little information about the transverse position of its trajectory. However, if the cells are staggered and only partially overlap with each other, it becomes more likely that the particle passed through the overlapping region. Exploiting this fact in the calorimeter's design by staggering cell positions in alternating layers improves its spatial resolution. The same principle applies to particle showers, provided that the cell size is comparable to the calorimeter's Moliere radius.

Staggering detector elements is a common technique in calorimeter design aimed at reducing dead areas and minimizing channeling effects. For instance, in the CMS's HGCAL design~\cite{CMSHGCAL}, the endcap petals are staggered with azimuthal rotations within three groups of layers to eliminate dead areas between petals~\cite{Noy_2022}. Additionally, in the ATLAS TileCal design~\cite{ATLASTileCalTDR}, quadrangular scintillator cells are positioned perpendicular to the beamline and staggered in the longitudinal direction.

In specific instances, staggering has been deliberately incorporated into detector designs to enhance spatial resolution. For example, the CALICE silicon-tungsten ECAL~\cite{CALICEcollaboration_2008,ILDConceptGroup:2020sfq} employs staggering of square cells to improve spatial resolution, as shown by a test beam study~\cite{CaliceStagger}. On the other hand, the CALICE tungsten-scintillator ECAL~\cite{CALICE:ScECAL_1,CALICE:ScECAL_2} employs a configuration of alternating vertical and horizontal strips.

While staggering alone can enhance the performance of traditional shower reconstruction algorithms, it also opens up possibilities for specialized algorithms that exploit the cell overlap. For instance, the Strip-Splitting Algorithm (SSA)~\cite{KOTERA2015158} was developed for calorimeters featuring alternating vertical and horizontal strips. This algorithm partitions each strip into sub-cells, to which energy is allocated based on the energy distribution in adjacent layers. As a result, the performance achieved with strips closely approximates what would be attainable with square pixels of dimensions equivalent to the strip widths. 

In this study, we advance the concept of staggering layers within high granularity calorimeters to enhance spatial resolution. Specifically, we introduce a novel staggering design based on a multifold repeating tessellation patterns. Additionally, we introduce the HEXPLIT algorithm, which harnesses the overlap between cells, akin to the way SSA utilizes strip layouts. What sets HEXPLIT apart from SSA is its use of overlaps between multiple layers, including adjacent and next-to-neighboring layers, each featuring distinct layouts. As a result, HEXPLIT can capitalize on multifold repeating patterns, yielding elevated levels of ``effective granularity''.

This paper is organized as follows: we present multifold repeating tessellation patterns in Section~\ref{sec:tessellations}, outline our simulation framework in Section~\ref{sec:simulations}, introduce the HEXSPLIT algorithm in Section~\ref{sec:reconstruction}, discuss our findings in Section~\ref{sec:results}, and conclude in Section~\ref{sec:conclusions}.

\section{Tessellations}
\label{sec:tessellations}
The choice of cell shape for creating a tessellation of an area influences the detector performance and potential enhancements of staggering configurations. The area, $A$, of a cell determines the total number of channels and the detector granularity. The circumradius, $r_{c}$, or maximum distance between the center of the cell and the vertices, is relevant for the light yield and cell uniformity when using the SiPM-on-tile technology~\cite{deSilva:2020mak,Belloni:2021kcw}. The amount of dead area between cells is proportional to the perimeter, $p$, of the cells. The single-cell spatial resolution is related to the RMS displacement from the centroid of the cell, defined geometrically by:
\begin{equation}
\Delta x = \sqrt{\frac{\iint_A dx\,dy\,x^2}{\iint_A dx\,dy}};\hspace{1cm}
\Delta y = \sqrt{\frac{\iint_A dx\,dy\,y^2}{\iint_A dx\,dy}}.
\label{eq:dxdy}
\end{equation}
For squares and rectangular strips, this yields the well-known result of $1/\sqrt{12}$ times the width or length. It should be noted that the resolution for showering particles can surpass the value given in Eq.~\ref{eq:dxdy} due to energy being distributed across cells. Nevertheless, this remains a valuable benchmark.

Table~\ref{tab:areas_and_rms} presents a summary of various properties pertaining to tessellating cell shapes that hold significance in calorimetry. It quantifies the well-known fact that hexagons possess the smallest perimeter and the smallest circumradius among all tessellating polygons for a given area. Moreover, hexagons have the highest count of symmetry axes, thereby enabling intricate staggering patterns.
\begin{table}[h!]
    \centering
    \begin{tabular}{c|c|c|c|c}
           & reg.~triangle & square & reg.~hexagon \\
        \hline
        $A$ & $\frac{\sqrt{3}}{4}s^2\approx 0.433s^2$ & $s^2$ & $\frac{3\sqrt{3}}{2}s^2\approx 2.60 s^2$  \\
        $r_c$ & $\frac{2}{3^{3/4}}\sqrt{A}\approx 0.877\sqrt{A}$& $\frac{1}{\sqrt{2}}\sqrt{A}\approx 0.707\sqrt{A}$& $\sqrt{\frac{2}{3\sqrt{3}}}\sqrt{A}\approx 0.620\sqrt{A}$ \\
        $p$ & $\frac{6}{\sqrt{\sqrt{3}}}\sqrt{A}\approx 4.46\sqrt{A}$& $4\sqrt{A}$& $6\sqrt{\frac{2}{3\sqrt{3}}}\sqrt{A}\approx 3.72\sqrt{A}$\\
        \multirow{ 2}{*}{$\Delta x$} & $\frac{1}{\sqrt{24}}s\approx0.204 s$ & $\frac{1}{\sqrt{12}}s\approx0.289s$  & $\sqrt{\frac{5}{24}}s\approx 0.456s$   \\
        & $\approx 0.310\sqrt{A}$ & $\approx0.289\sqrt{A}$ & $\approx 0.283\sqrt{A}$ \\
        $\Delta y$ & same as $\Delta x$ & same as $\Delta x$ & same as $\Delta x$ 
    \end{tabular}
    \caption{Summary of the properties of tessellating regular polygons, including the area ($A$), side length ($s$), circumradius ($r_c$), perimeter ($p$), and the RMS displacement in the $x$ ($\Delta x$) and $y$ ($\Delta y$) directions.}
    \label{tab:areas_and_rms}
\end{table}

\begin{figure}[h!]
     \centering
              \includegraphics[width=0.66\textwidth]{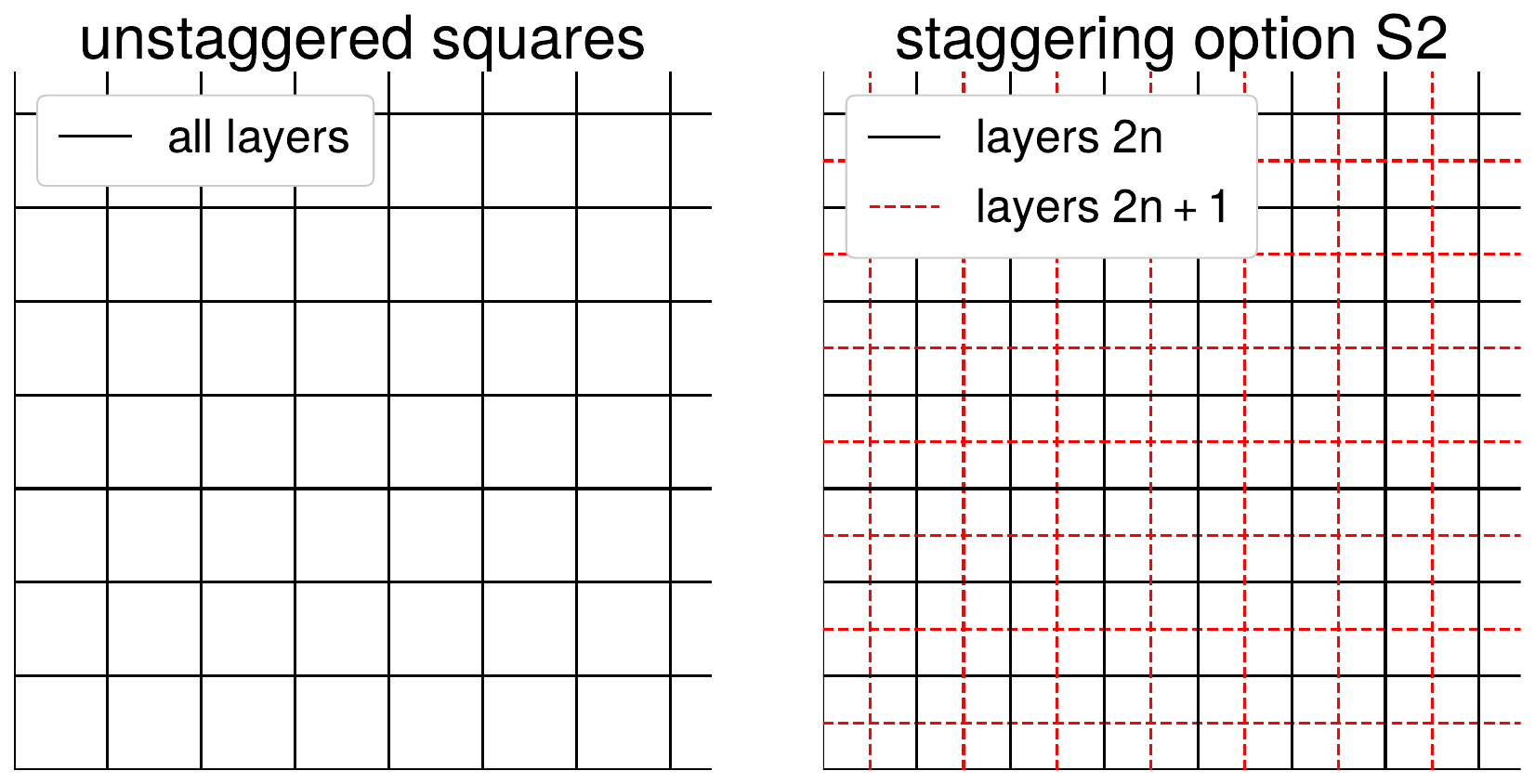}
     \includegraphics[width=1.0\textwidth]{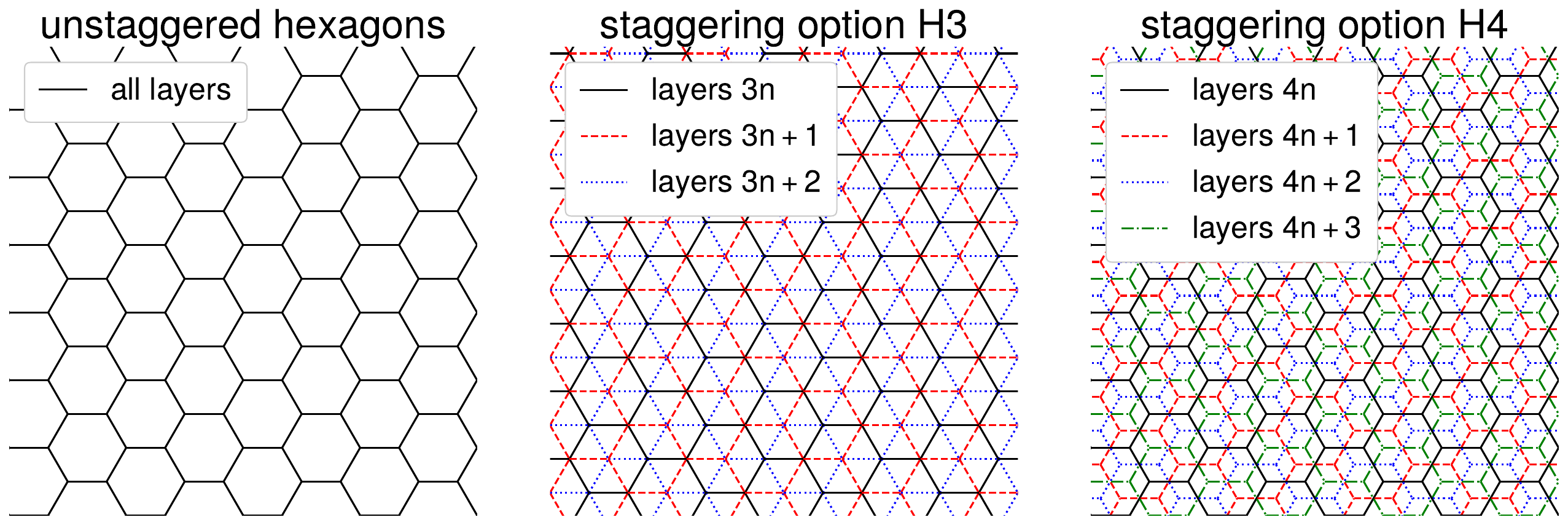}
              \caption{Top row, from left to right: unstaggered squares, and the S2 iteration of the square tessellation. Bottom row, from left to right: non-staggered hexagons, the H3 iteration of the hexagonal tessellation, and the H4 iteration of the hexagonal tessellation.}
     \label{fig:tessellations}
 \end{figure}

Figure~\ref{fig:tessellations} displays various tessellations, both square- and hexagon-based, with and without staggering. The top row illustrates a square tessellation without staggering on the left and with a simple\footnote{In the CALICE ECAL beam test, a more intricate staggering pattern was employed along one axis~\cite{CALICEcollaboration_2008}.} staggering pattern on the right, denoted as S2. Additionally, we introduce two novel staggered tessellation patterns with hexagonal cells, labeled as H3 and H4. These two staggering configurations repeat every three and four layers, respectively.

In H3, the centers of the cells on one layer have the same transverse position as the corners of the cells on the other layers. Equivalently, the centers of the cells in subsequent layers are one hexagon radius apart. This produces triangular subcells, each of which is 1/6 of the area of the hexagonal cells and has $\Delta x_{\rm tri}=\Delta y_{\rm tri}\approx 0.204 s_{\rm hex}$, where $s_{\rm hex}$ is the side length of the hexagons.

In H4, the centers of the cells on one layer are at the midpoints of the edges of the cells in the other layers. Equivalently, the cell centers are shifted by $\sqrt{3}/2$ hexagon radii from one layer to another. Effectively, groups of four layers define rhombus-shaped subcells, each of which is 1/12 of the area of the hexagonal cells. For a rhombus, the RMS displacement from the centroid of the cell is $\frac{1}{\sqrt{8}}s\approx0.354 s$ in the long direction and $\frac{1}{\sqrt{24}}s\approx 0.204 s$ in the short direction, where $s$ is the length of one side of the rhombus. However, since the side length of the rhombical sub-cells is half of that of a hexagonal cell, we have $\Delta x_{\rm rhombus}\approx 0.177 s_{\rm hex}$ and $\Delta y_{\rm rhombus}\approx 0.102 s_{\rm hex}$.

All of these tessellation patterns could be practically implemented in realistic detector designs. For example, they could be seamlessly integrated into the SiPM-on-tile approach, where extensive layers of scintillator cells are tessellated to span a large-area surface. This integration could be achieved by encasing individual cells in reflective foil, akin to the CALICE AHCAL~\cite{CALICE:2022uwn} and CMS HGCAL~\cite{CMSHGCAL} configurations. Additional techniques, like attaching individual cells onto plastic frames~\cite{Arratia:2023rdo} to establish unique patterns, could also present viable alternatives.
\section{Simulations}
\label{sec:simulations}
We employed the \textsc{DD4HEP}~\cite{Frank:2014zya} framework, which internally relies on \textsc{GEANT4}~\cite{GEANT4:2002zbu} with the FTFP\_BERT\_HP physics list, to simulate the geometry\footnote{We adopted a coordinate system aligned with the calorimeter, wherein the $z$-axis is perpendicular to the layers, $y$ represents the vertical direction, and $x$ is orthogonal to both $y$ and $z$, establishing a right-handed coordinate system.} of an iron-scintillator calorimeter akin to the CALICE AHCAL~\cite{CALICE:2022uwn}, CMS HGCAL~\cite{CMSHGCAL}, and the ePIC forward calorimeter insert~\cite{hcalInsert}. This setup comprises alternating layers of 20 mm-thick iron absorbers and layers of 3 mm-thick scintillator tiles. In total, 64 layers were included, for a total of 8.0$\lambda$. The detector area is 6$\times$6 m$^2$. 

Each cell is represented as either a hexagonal or square shape, both with an area of 25 cm$^2$. The circumradius of the cells is approximately 1.0 or 1.1 times the Molière radius of this detector (3.1 cm), respectively for the hexagonal and square shapes. A repeating tessellation pattern using either hexagons or square cells is implemented. Since \textsc{DD4HEP} does not include built-in code for hexagonal segmentation, we developed our own plugin for this purpose~\cite{stagseg}.

We simulated 2,000 single-neutron events within the energy range of 10--300 GeV for each configuration (unstaggered squares, S2, unstaggered hexagons, H3, and H4), covering polar angles from 0 to 5.5 mrad to avoid potential artifacts that might arise from constraining the angle to a single value. The part of the detector's front face impacted by neutrons encompassed an area of approximately 1200 cm$^2$. Furthermore, we conducted additional simulations involving muons and electrons to determine the hit-energy MIP unit and establish the calorimeter's electromagnetic energy scale. The MIP unit, defined as the most probable value, was found to be MIP$=0.5$ MeV.

Hits are reconstructed using a 15-bit ADC that is free from electronic noise, as well as optical and electronic cross-talk. In the subsequent analysis, only hits with $E > 0.1$ MIP and $t < 200$ ns are taken into account. All hits surpassing these thresholds are retained for further analysis, without applying any clustering algorithms.

\section{Shower-Reconstruction Algorithm}
\label{sec:reconstruction}
As a baseline, we evaluate the single-particle position resolution using a conventional procedure, as described in Section~\ref{sec:log_weight}. Subsequently, we introduce HEXPLIT, an algorithm that exploits sub-cell information, outlined in Section~\ref{sec:subcell}.

\subsection{Log-weighted determination of the shower position}
\label{sec:log_weight}
The position of the shower is reconstructed using a logarithmically weighted average of the center positions of the cells~\footnote{We also experimented with both the baseline and HEXPLIT algorithms by employing weights linearly proportional to energy in neutron showers. However, this approach yielded significantly poorer resolutions compared to the logarithmic weighting discussed here.}, similar to Ref.~\cite{Acar_2022, Akchurin_2018}.  We used the following formula:
\begin{equation}
    \vec x_{\rm recon}=\frac{\sum\limits_{i\in \rm hits}\vec x_iw_i}{\sum\limits_{i\in\rm hits}w_i}
    \label{eq:xrecon_hits}
\end{equation} 
where $\vec x_i$ are the 3d positions of the hits, and the weights, $w_i$, are determined by 
\begin{equation}
    w_i=\max \left(0, w_0+\ln{\frac{E_i}{E_{\rm tot}}}\right)
    \label{eq:weight}
\end{equation}
Here, $E_i$ represents the energy of an individual hit, and $E_{\rm tot}$ denotes the total energy of the shower. The parameter $w_0$ governs the minimum hit energy used. The optimal value for $w_0$ varies depending on energy and hardware parameters, such as cell size and the Molière radius of the detector.  
\subsection{Sub-cell reweighting with HEXPLIT}
\label{sec:subcell}
The HEXPLIT algorithm involves the energy reweighting of sub-cells formed through the overlap of cells within a specific layer and neighboring layers. The initial step involves characterizing the energy distribution within these sub-cells. Following that, this process uses the positions and energies of these sub-cells as input for the logarithmic weighting procedure outlined in Section~\ref{sec:log_weight}.

Assigning a weight to each sub-cell in a staggering cycle, denoted by the number of layouts $N$, is accomplished using the following formula:
\begin{equation}
W_i = \prod\limits_{j=1}^{N-1}{\rm max}(E_{j},\delta),
\label{eq:Wi_analytical}
\end{equation}
Here, $i$ denotes the index of a subcell.  The product over $j$ is over the tiles in the nearby layers that overlap with subcell $i$, and $E_j$ represents the energy in that tile.  The parameter $\delta$ acts as a noise threshold and prevents division by zero. For this analysis, we set $\delta$ to 1 MIP.

In scenarios where $N$ is even, such as in the instances of H4 and S2, the layouts on layers that are $N/2$ layers distant from the current layer exhibit identical patterns. Consequently, we introduce a virtual tile, whose energy is the aggregate of the energies from the tiles within those layers overlapping the $i^{\rm th}$ subcell.

In the H3 configuration, this is analogous to taking a product over the overlapping tiles in the layer immediately preceding and following the $i^{\rm th}$ subcell. In H4, the calculation encompasses both the layers immediately preceding and following, as well as the virtual layer generated by summing the energies from two layers downstream and two layers upstream of the $i^{\rm th}$ subcell. In the S2 arrangement, only the virtual layer computed from the summation of energies in the directly preceding and following layers is used.

We proceed to assign the updated signal within the current subcell as follows:
\begin{equation}
E_i = E_{\rm tile} W_i/\sum_j W_j.
\end{equation}
Here, the summation of weights across $j$ serves to normalize the weights, ensuring that the total signal within the tile remains unchanged, i.e., $\sum_i E_i = E_{\rm tile}$.

The shower's position is subsequently calculated in a manner analogous to Eq.~\ref{eq:xrecon_hits}:
\begin{equation}
    \vec x_{\rm recon}=\frac{\sum\limits_{i\in \rm subcells}\vec x_iw_i}{\sum\limits_{i\in\rm subcells}w_i}
    \label{eq:xrecon_subcells}
\end{equation} 
where $\vec x_i$ are the 3d positions of the subcells, and the weights, $w_i$, are determined by 
\begin{equation}
    w_i=\max \left(0, w_0+\ln{\frac{E_i}{E_{\rm tot}}}\right)
    \label{eq:weight_subcells}
\end{equation}
where $E_i$ is the energy of the subcell, and $E_{\rm tot}$ is the total energy of the shower.  

This technique extends the principles of the SSA method~\cite{KOTERA2015158}, which focused on the overlap between vertical and horizontal strips. A notable distinction between HEXPLIT and the SSA lies in their applicability. The SSA is tailored for strip geometries in which the layers directly preceding and following a given layer share identical layouts, \textit{e.g.} those used in Ref.\cite{CALICE:ScECAL_1,CALICE:ScECAL_2,CEPC_ECAL,FDC}. On the contrary, HEXPLIT extends this concept to encompass non-strip geometries and multifold tessellating cycles. In this framework, the intersection of multiple adjacent layers, each characterized by a distinct pattern, plays a role in establishing the subcell boundaries and their associated energies.

For detectors where the cell size varies per layer (as is being considered for the EIC calorimeter insert design~\cite{hcalInsert}), the weights in Eq.~\ref{eq:weight_subcells} should be adjusted by a factor that is inversely proportional to the area of an individual cell.

\subsection{Example reconstruction of a hadronic shower}
\begin{figure}[hb!]
    \centering
    \includegraphics[width=0.65\textwidth]{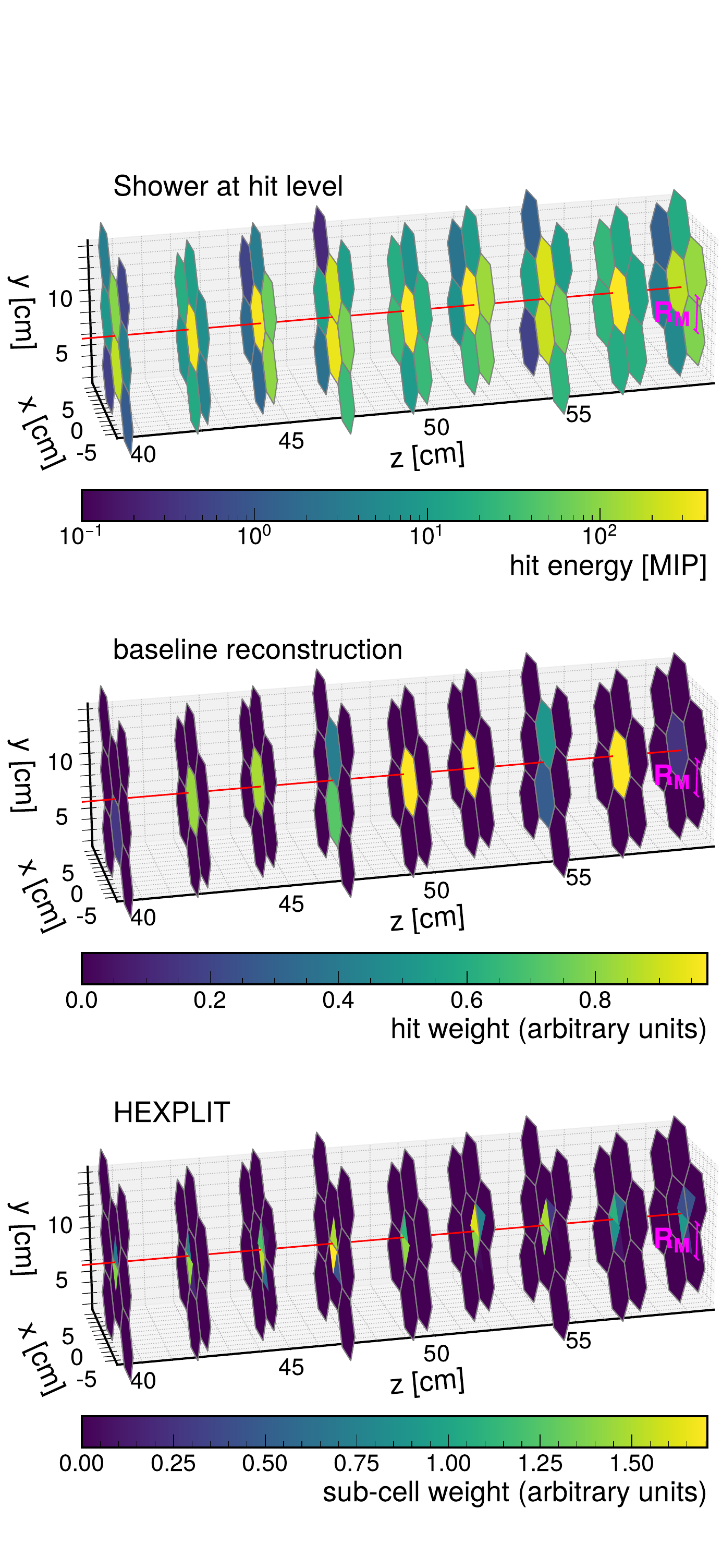}
     \caption{
Top: A segment of a simulated neutron shower example employing the H3 staggered configuration.
Middle: Weights of the hits in the baseline shower-position reconstruction algorithm.
Bottom: Weights of the sub-cells in the shower-position reconstruction using the HEXPLIT algorithm, incorporating sub-cell reweighting. The red line depicts the trajectory of the neutron, extrapolated from its generator-level position and angle. In this coordinate system, $z=0$ corresponds to the front face of the detector.
 }
    \label{fig:algorithm_h3}
\end{figure}

To illustrate the improvement in positional resolution achieved with HEXPLIT, we present a section of a simulated neutron shower measured using the H3 configuration in Fig.~\ref{fig:algorithm_h3}. The top panel displays the energy distribution across individual cells. In the middle panel, we demonstrate the shower-position reconstruction using the conventional log-weighting method. In this representation, only cells that contribute to position determination are shown with non-zero values. The color spectrum corresponds to the weights used in determining the shower position, with a threshold of $w_0=4.0$ applied. The bottom panel shows the results when using HEXPLIT. The colors assigned to the triangular sub-cells represent the weights attributed to the sub-cell contributions in determining the shower position.

\FloatBarrier
\section{Results}
\label{sec:results}
\begin{figure}[h!]
    \centering
    \includegraphics[width=\textwidth]{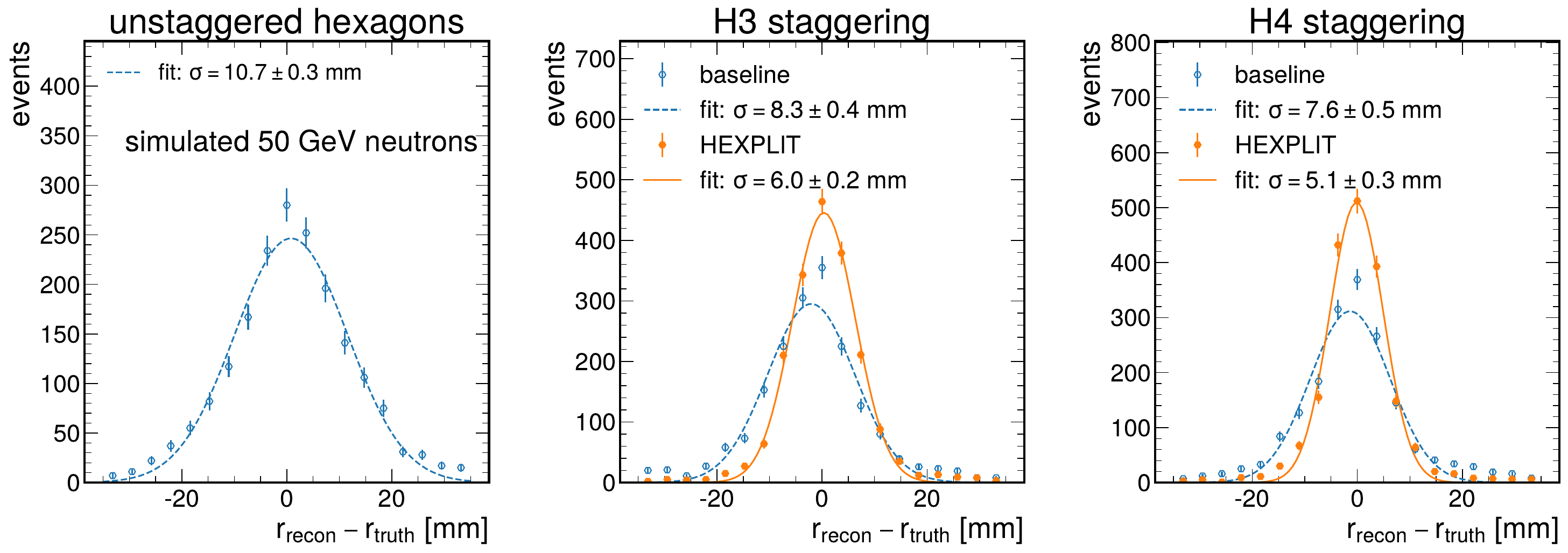}
    \includegraphics[width=0.67\textwidth]{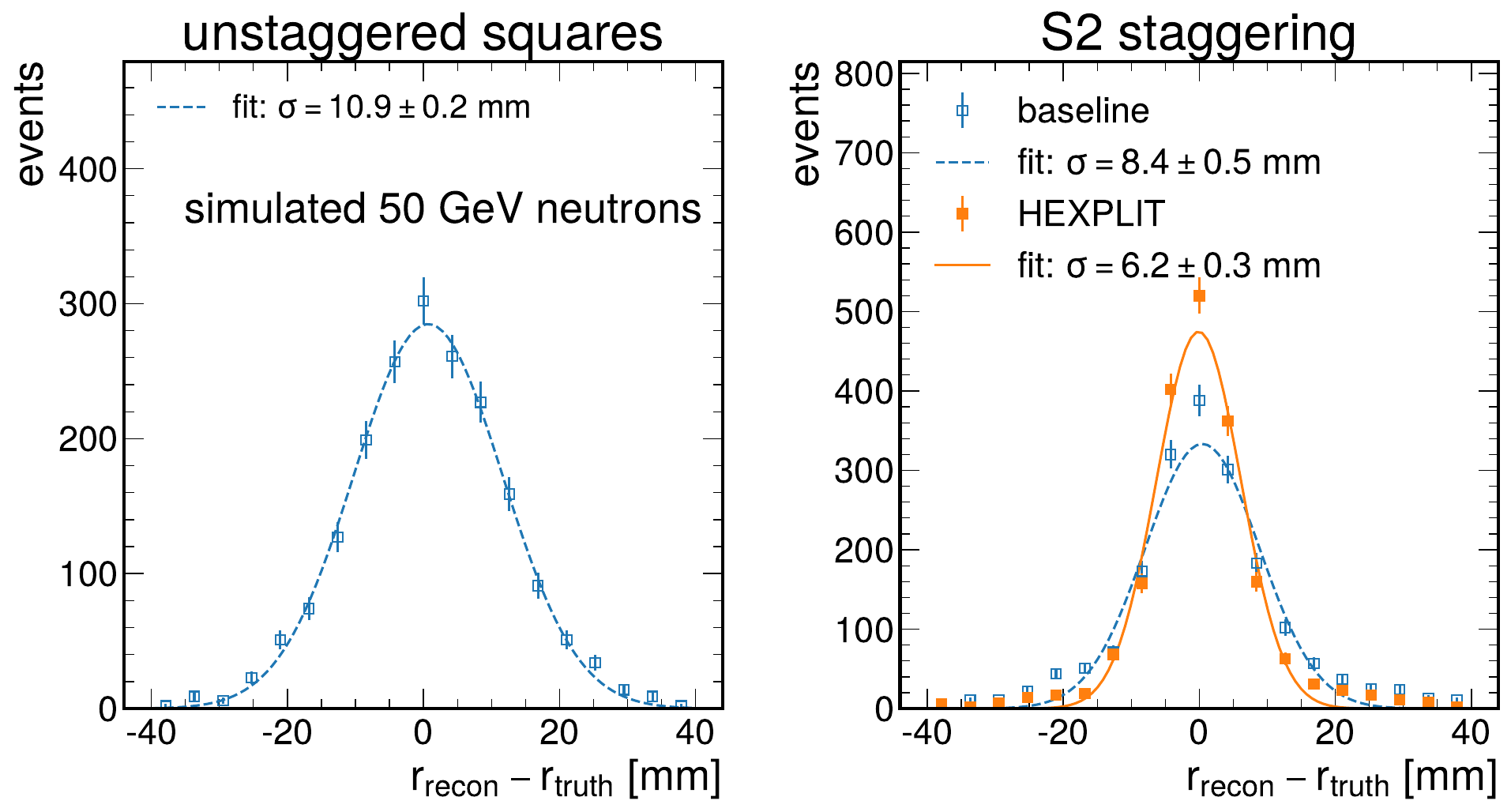}
    \caption{Residual distributions of neutron-shower position reconstruction.  Top row: results with the unstaggered hexagons configuration (left), staggered configuration H3 (middle) and the staggered configuration H4 (right). 
 Bottom row: results with the unstaggered squares configuration (left), and the staggered configuration S2 (right). The results for the staggered configurations are shown separately with the baseline reconstruction (blue, open symbols) and with the HEXPLIT algorithm (orange, filled symbols).  Each of these distributions are fit to Gaussian functions (curves).  }
    \label{fig:fits}
\end{figure}

Figure~\ref{fig:fits} shows the  distributions of the radial-position residuals, defined by 
\begin{equation}
    \Delta r=r_{\rm recon}-r_{\rm truth},
\end{equation}
for 50 GeV neutrons in each of hexagonal-cell and square-cell layouts described in Sec.~\ref{sec:tessellations}.  
Here, $r_{\text{recon}}$ is the reconstructed radial position of the shower, while $r_{\text{truth}}$ represents the true position of the incoming particle.
More specifically, the value of $r_{\rm recon}$ is given by $\sqrt{x_{\rm recon}^2+y_{\rm recon}^2}$, where $x_{\rm recon}$ and $y_{\rm recon}$ are given by Eq.~\ref{eq:xrecon_hits} for the baseline algorithm and Eq.~\ref{eq:xrecon_subcells} for HEXPLIT.

The value of $r_{\text{truth}}$ is analogously defined for the point along the truth-particle trajectory that aligns with the same longitudinal position as the reconstructed position (i.e., the log-weighted center of the shower, as represented by the $z$ component in Eq.~\ref{eq:xrecon_hits} or Eq.~\ref{eq:xrecon_subcells}). This formulation of position residuals mirrors the approach used in Ref.~\cite{Acar_2022}.

For the $w_0$ parameter, we used the following functional form
\begin{equation}
    w_0(E)=a+b\,\log \frac{E}{\rm 50\,GeV}+c\,(\log \frac{E}{\rm 50\,GeV})^2,
\end{equation}
Where $E$ represents the reconstructed energy in GeV, and $a$ and $b$ are constants. The values of $a$, $b$, and $c$ used in this study are presented in Table~\ref{tab:ab}. These values were determined through inspection to yield optimal outcomes.
\begin{table}[]
    \centering
    \begin{tabular}{c|ccc}
        configuration & $a$ & $b$ & $c$\\
        \hline
         unstaggered hexagons & 5.9  & 0.33 & 0.0 \\
         H3 baseline & 4.0 & 0.019 & 0.59 \\
         H4 baseline & 4.2 & 0.045 & 0.32 \\
         H3 HEXPLIT & 5.0 & 0.89 & 0.17 \\
         H4 HEXPLIT & 5.0 & 0.65 & 0.31 \\
    \end{tabular}
    \caption{Values of $a$, $b$, and $c$ used to calculate the $w_0$ value used in reconstruction.}
    \label{tab:ab}
\end{table}

In the top row of Fig.~\ref{fig:fits}, we present the residual distributions obtained for 50 GeV neutrons for five distinct scenarios: unstaggered hexagons, H3 staggering with the baseline reconstruction algorithm, H4 staggering with the baseline reconstruction algorithm, H3 staggering with HEXPLIT, and H4 staggering with HEXPLIT.

To derive the resolutions, we did fit each distribution with a Gaussian function and extracted the $\sigma$ values from these fits. Without the implementation of staggering, the resolution measured at 50 GeV is 10.7$\pm$0.3 mm. By employing H3 staggering, we achieved an improved resolution of 8.3$\pm$0.4 mm using the baseline reconstruction, which further improved to 6.0$\pm$0.2 mm with HEXPLIT.

For configurations involving H4 staggering, the results were somewhat better than those of H3 and unstaggered in the context of baseline reconstruction, resulting in a resolution of 7.6$\pm$0.5 mm. With the integration of HEXPLIT, the resolution advanced significantly to 5.1$\pm$0.3 mm.

For completeness, the results for 50 GeV neutrons with square tessellations—both staggered and unstaggered—are shown in the bottom row of Fig.~\ref{fig:fits}. Unstaggered squares resulted in a resolution of 10.9$\pm$0.2 mm. For staggered squares using baseline reconstruction, the resolution was slightly better but similar, at 8.4$\pm$0.5 mm. On the other hand, staggered squares employing HEXPLIT reconstruction achieved a resolution of 6.2$\pm$0.3 mm, representing a 43\% improvement over unstaggered squares. The optimal resolutions were attained by using specific $w_0$ values for these square-cell configurations: 6.3 for unstaggered squares, 4.5 for staggered squares with baseline reconstruction, and 5.0 for staggered squares with HEXPLIT reconstruction.

\begin{figure}[h!]
     \centering
     \includegraphics[width=0.6\textwidth]{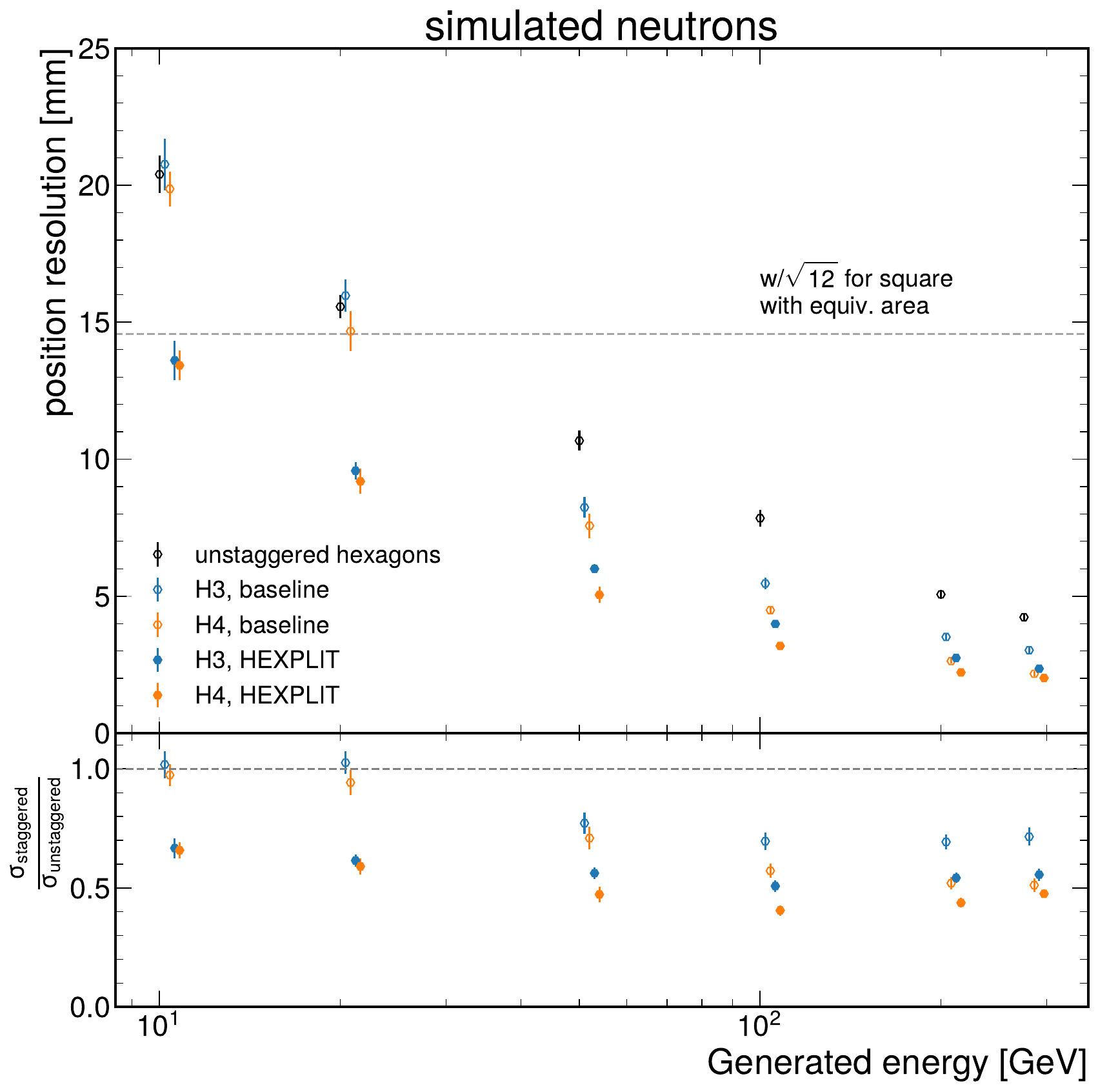}
     \caption{Top Panel:
Neutron-shower position resolution comparisons among different configurations, including the unstaggered hexagon configuration (black), H3 staggered configuration (blue), and the H4 staggered configuration (orange). For the staggered simulations, both resolutions obtained with (solid symbols) and without (open symbols) HEXPLIT are displayed. The horizontal line represents the resolution anticipated from the side length divided by $\sqrt{12}$ for squares with the same area as the hexagons used in the simulations.
Bottom Panel: Ratio depicting the resolutions in the staggered hexagonal configurations relative to those of the unstaggered hexagon configuration.
}
     \label{fig:position_resolution}
 \end{figure}

In the top panel of Fig.~\ref{fig:position_resolution}, we present the position resolution as a function of neutron energy and compare the results between the unstaggered geometry and the H3 and H4 staggered configurations. For the staggered layouts, a comparison is made between results obtained with the baseline reconstruction and those achieved with HEXPLIT. The bottom panel depicts the extent of improvement achieved by the staggered layouts when compared to the unstaggered arrangement.

At the lowest energies, the unstaggered and both staggered layouts have about the same resolutions when employing the baseline algorithm. With the application of the HEXPLIT algorithm, there is about a 30\% improvement for both staggered layouts. As the neutron energies increase, the H4 staggering outperforms the H3 staggering, which in turn exhibits better resolution than the unstaggered hexagonal configuration.

For mid-to-high energies, the H4 staggering with the baseline reconstruction results in a substantial improvement, up to 50\% (depending on the energy) over the unstaggered configuration. With the HEXPLIT algorithm, the relative improvement of H4 in comparison to the unstaggered configuration is even more pronounced, from about 35\% to 60\% across the full energy range.

To investigate the feasibility of the staggered layouts and the HEXPLIT algorithm in electromagnetic calorimetry, we repeated our studies for single-photon events, as described in more detail in Appendix~\ref{sec:photons}. We found that the position resolutions obtained for photon events with the baseline reconstruction improved substantially with both the H3 and H4 staggering layouts. However, the HEXPLIT algorithm did not further enhance the performance as it did for neutron showers. Overall, the resolutions were improved twofold through the use of H4, similar to the neutron results.

\begin{figure}
    \centering
    \includegraphics[width=0.6\textwidth]{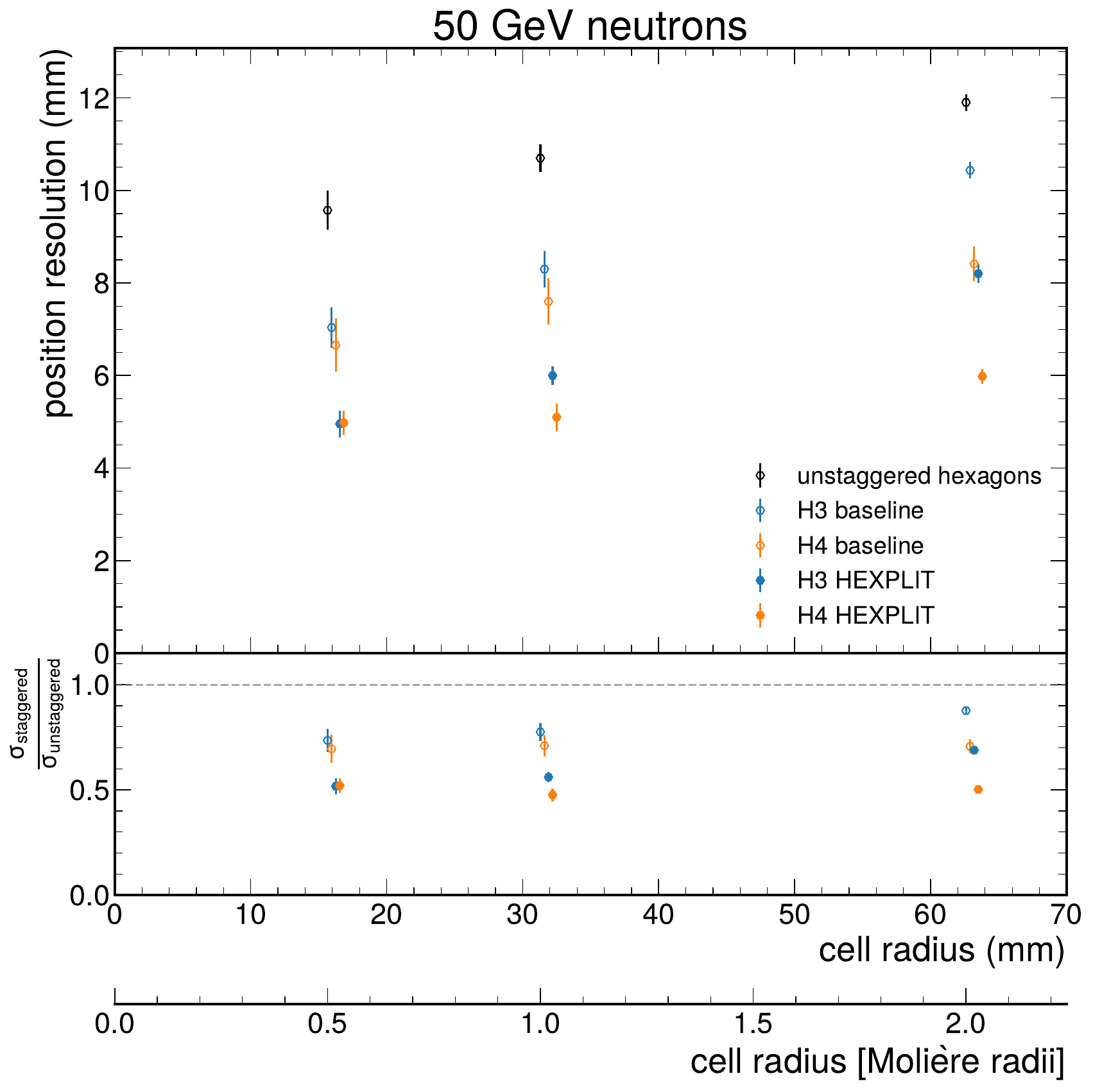}
    \caption{Neutron-shower position resolution comparisons among different configurations and cell sizes, including the unstaggered hexagon configuration (black), H3 staggered configuration (blue), and the H4 staggered configuration (orange). For the staggered simulations, both resolutions obtained with (solid symbols) and without (open symbols) HEXPLIT are displayed. 
Bottom Panel: Ratio depicting the resolutions in the staggered hexagonal configurations relative to those of the unstaggered hexagon configuration. }
    \label{fig:cell_sizes}
\end{figure}

We explored the impact of cell size on resolution by conducting 50 GeV neutron simulations with hexagons that were either twice as large or half as large as the original cells. The resolutions for all three sizes are presented in Fig.~\ref{fig:cell_sizes}. Our findings indicate that in every case, staggered designs outperform the unstaggered design. Additionally, H4 staggering produces better outcomes than H3, and both configurations show improvement when HEXPLIT is applied. We also observed that resolution improvement reaches a point of diminishing returns as the cell sizes decrease.

\section{Summary and Conclusions}
\label{sec:conclusions}
We have developed an algorithm and introduced innovative tessellation patterns that significantly enhance position resolution for hadronic showers detected by high-granularity calorimeters. Our findings underscore the performance improvements achieved by incorporating staggered hexagonal cells in cyclic tessellating patterns, resulting in enhancements of up to nearly double the performance compared to non-staggered configurations.

These findings highlight the substantial potential of staggered configurations, which can also be harnessed in machine-learning applications~\cite{Paganini:2017dwg,Paganini:2017hrr,Belayneh:2019vyx,Qasim:2019otl,DiBello:2020bas,Buhmann:2020pmy,Akchurin:2021afn,Pata:2021oez,Neubuser:2021uui,Akchurin:2021ahx,Buhmann:2021caf,Khattak:2021ndw,Chadeeva:2022kay,Qasim:2022rww,Mikuni:2022xry}, particularly those capable of accommodating complex geometries~\cite{Qasim:2019otl,Buhmann:2023bwk,Liu:2023lnn,Acosta:2023zik,Amram:2023onf,Hashemi:2023ruu}.

The improved position resolution has the potential to enhance the performance of the particle-flow algorithm~\cite{Thomson_2009} for a given detector granularity. Alternatively, this technique can be employed to optimize costs by reducing the necessary number of channels in the design of high-granularity calorimeters, all while maintaining the desired level of spatial resolution.

These outcomes pave the way for future explorations into the viability of employing staggered scintillator tiling in other ongoing high-granularity sampling calorimeter projects, including those at the HL-LHC~\cite{CMSHGCAL}, ILC~\cite{ILCTDR}, CEPC~\cite{CEPCStudyGroup:2018ghi}, and the EIC~\cite{hcalInsert,Bock:2022lwp}. 
\section{Code Availability}
An implementation of the HEXPLIT algorithm, along with example simulated files, can be found in Ref.~\cite{hexplitExamples}.
 \section*{Acknowledgments}
We thank Sean Preins for insightful discussions on tessellation with hexagons in the design of the calorimeter insert. We thank members of the California EIC consortium for their feedback on our work. This work was supported by MRPI program of the University of California Office of the President, award number 00010100. S.P also acknowledges support from the Jefferson Lab EIC Center Fellowship. M.A acknowledges support through DOE Contract No. DE-AC05-06OR23177 under which Jefferson Science Associates, LLC operates the Thomas Jefferson National Accelerator Facility.
\bibliographystyle{utphys} 
\bibliography{biblio.bib}
\appendix
\section{Electromagnetic calorimetry}
\label{sec:photons}
\begin{figure}
    \centering
    \includegraphics[width=\textwidth]{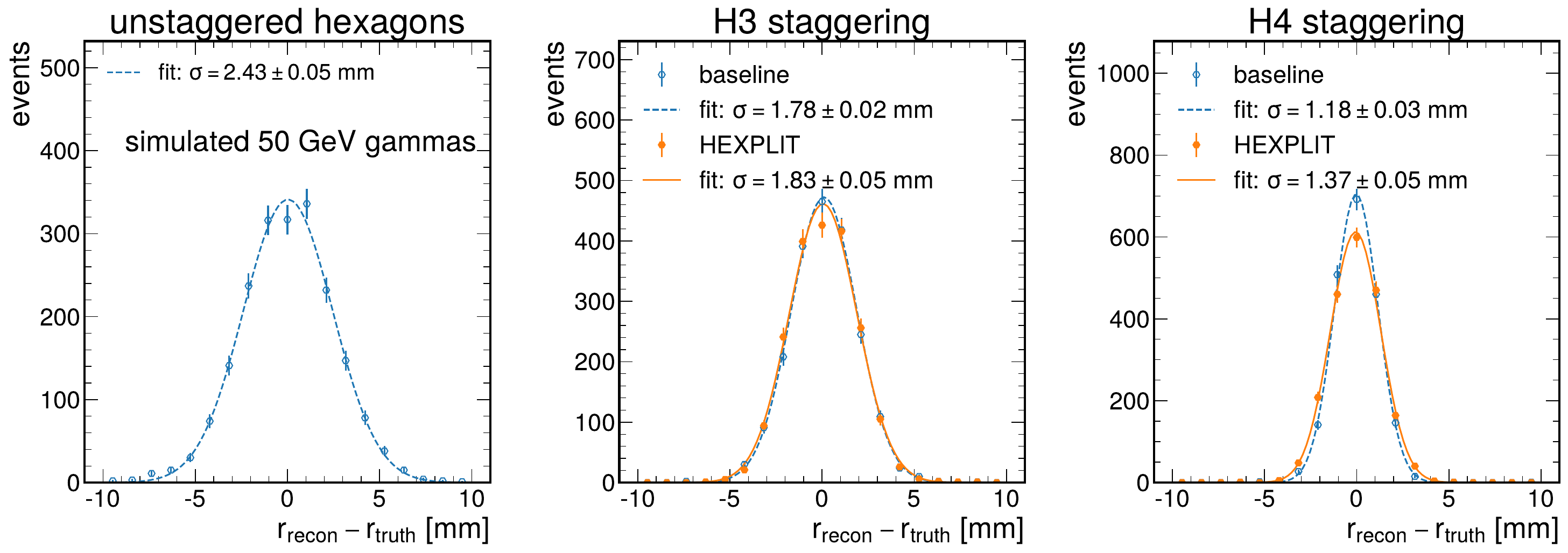}
    \caption{Residual distributions of photon-shower position reconstruction. Results with the unstaggered hexagons configuration (left), staggered configuration H3 (middle) and the staggered configuration H4 (right). The results for the staggered configurations are shown separately with the baseline reconstruction (blue, open symbols) and with the HEXPLIT algorithm (orange, filled symbols).  Each of these distributions are fit to Gaussian functions (curves). }
    \label{fig:fits_photon}
\end{figure}

We repeated our studies with photons instead of neutrons. We used a separate optimization of the $w_0$ parameters for this study, using the parameters in Table~\ref{tab:ab_photon}.  We show the distributions of the position residuals for 50 GeV photons in Fig.~\ref{fig:fits_photon}, analogous to Fig.~\ref{fig:fits}.  We found that for each of the layouts considered (unstaggered hexagons, H3, and H4), the position resolutions obtained using the baseline reconstruction for photons are several times better than those for neutrons (see Fig.~\ref{fig:fits}). Furthermore, the staggering pattern H3 offers better resolution than the unstaggered hexagons, and the H4 staggering has better resolution than H3. However, for both the H3 and H4 staggering patterns, HEXPLIT is found to yield performance similar to the baseline algorithm. This may be a limitation of the detector geometry and the cell size considered.

\begin{figure}
    \centering
    \includegraphics[width=0.6\textwidth]{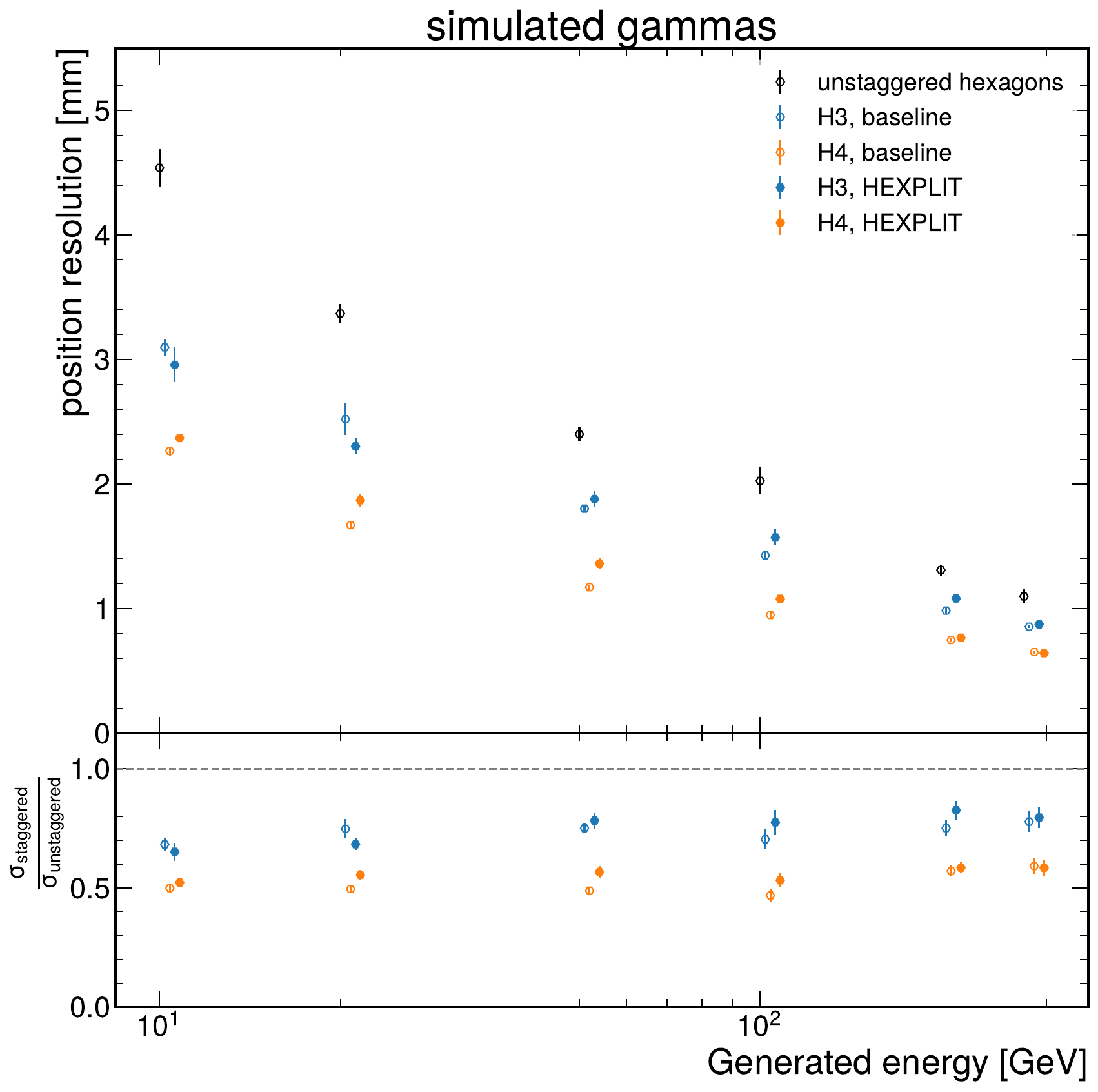}
    \caption{Photon-shower position resolution comparisons among different configurations, including the unstaggered hexagon configuration (black), H3 staggered configuration (blue), and the H4 staggered configuration (orange). For the staggered simulations, both resolutions obtained with (solid symbols) and without (open symbols) HEXPLIT are displayed. Bottom Panel: Ratio depicting the resolutions in the staggered hexagonal configurations relative to those of the unstaggered hexagon configuration. }
    \label{fig:position_resolution_photon}
\end{figure}

We show the dependence of the position resolution on the photon energy in Fig.~\ref{fig:position_resolution_photon}, analogous to Fig.~\ref{fig:position_resolution}.  The resolution with the H3 staggering using the baseline algorithm is around 30\% better than that of the unstaggered hexagons across the full energy range, while the H4 staggering is about 50\% better than the unstaggered hexagons.

\begin{table}[]
    \centering
    \begin{tabular}{c|ccc}
        configuration & $a$ & $b$ & $c$\\
        \hline
         unstaggered hexagons & 7.3  & $-$0.14 & $-$0.088 \\
         H3 baseline & 6.3 & 0.25 & $-$0.23 \\
         H4 baseline & 5.4 & 0.37 & 0.0 \\
         H3 HEXPLIT & 7.6 & 0.38 & 0.0 \\
         H4 HEXPLIT & 6.0 & -0.46 & 0.099 \\
    \end{tabular}
    \caption{Values of $a$, $b$, and $c$ used to calculate the $w_0$ value used in the reconstruction of photons.}
    \label{tab:ab_photon}
\end{table}

\end{document}